\documentclass[prd,twocolumn,superscriptaddress,altaffilletter,showpacs,nofootinbib]{revtex4}
\usepackage{amssymb}
\usepackage[dvips]{graphicx}
\usepackage{amsmath}

\newcommand{\be}{\begin{equation}}
\newcommand{\ee}{\end{equation}}
\newcommand{\bea}{\begin{eqnarray}}
\newcommand{\eea}{\end{eqnarray}}
\newcommand{\bib}{\bibitem}
\newcommand{\der}{\partial}
\newcommand{\vphi}{\varphi}

\begin{document}

\title{Scale invariance and broken electroweak symmetry may coexist together}

\author{Israel Quiros}\email{iquiros6403@gmail.com}\affiliation{Departamento de Matem\'aticas, Centro Universitario de Ciencias Ex\'actas e Ingenier\'{\i}as (CUCEI), Corregidora 500 S.R., Universidad de Guadalajara, 44420 Guadalajara, Jalisco, M\'exico.}

\date{\today}

\begin{abstract}
Here we show that local scale invariance -- invariance under Weyl rescalings -- may safely coexist with broken electroweak symmetry if assume the Weyl geometric theory to govern the affine structure of spacetime. We find that within the resulting scale invariant theory of gravity cosmological inflation is enough to explain the large hierarchy between the Higgs and Planck masses.
\end{abstract}

\pacs{04.50.Kd, 11.15.-q, 11.15.Ex, 98.80.Cq}

\maketitle

The apparently straightforward statements made by Dicke in Ref. \cite{dicke} about the naturalness of requiring invariance of the laws of physics under transformations of units, and about the fact that there may be more than one feasible way of establishing the equality of units at different spacetime points, raise the question about considering generalizations of (pseudo)Riemann geometry. The first such generalization that comes to one's mind is Weyl geometry \cite{weyl, adler, bib-weyl, novello}. Weyl's geometric theory is no more than a generalization of Riemann geometry to include point dependent length of vectors during parallel transport, in addition to the point dependent property of vectors directions. It is assumed that the length of a given vector ${\bf l}$ ($l\equiv\sqrt{g_{\mu\nu} l^\mu l^\nu}$) varies from point to point in spacetime according to: $dl=l w_\mu dx^\mu/2$, where $w_\mu$ is the Weyl gauge boson. Hence, if attach a spacetime vector to a given standard unit of length, the second of the Dicke's statements above finds a natural realization in Weyl geometry. The first of the statements made by Dicke can be implemented in any theory of gravity which is invariant under the Weyl rescalings/local scale transformations \cite{deser, waldron}:

\bea g_{\mu\nu}\rightarrow \Omega^2g_{\mu\nu},\;w_\mu\rightarrow w_\mu-2\der_\mu\ln\Omega,\label{weyl-r}\eea where the (smooth) positive spacetime function $\Omega^2=\Omega^2({\bf x})$ is the conformal factor, and the spacetime coincidences/coordinates are kept unchanged. Regrettably, according to the most widespread understanding of scale transformations (\ref{weyl-r}), invariance under the Weyl rescalings is dynamically broken through the Higgs mechanism and, in the presence of matter, invariance under the transformations of units in Dicke's sense seems to be forbidden.\footnote{The conformal transformation of the metric in (\ref{weyl-r}) is what Dicke regards as a transformation of units in \cite{dicke}.} 

In this letter we shall show that, as a matter of fact, scale invariance and symmetry breaking can be compatible concepts, i. e., local scale invariance and broken electroweak (EW) symmetry may safely coexist together. The consequence of this for particle physics and cosmology is very interesting: nothing besides a consistent scale-invariant theory of gravity and particles, and cosmological inflation is necessary to explain the large hierarchy between the Planck and EW energy scales.

\subsection*{Scale invariance and EW symmetry breaking}

A typical argument against invariance under Weyl rescalings goes like this. Consider the following modification of the proposal \cite{deser, dirac} by replacing the scalar field by the Higgs field of the standard model of particles (SMP) \cite{cheng}:

\bea &&S=\int d^4x\sqrt{|g|}\left[\frac{\xi|h|^2}{2}\,R^{(w)}-\right.\nonumber\\
&&\left.\;\;\;\;\;\;\;\;\;\;\;\;\;\;\;g^{\mu\nu}(D_\mu h)^\dag(D_\nu h)-\frac{\lambda}{4}(|h|^2-v_0^2)^2\right],\label{cheng-action}\eea where $\xi$ is the non minimal coupling parameter, we are using the unitary gauge for the Higgs isodoublet $H^T=(0,h)/\sqrt{2}$, and $|h|^2\equiv h^\dag h$. In equation (\ref{cheng-action}) $R^{(w)}$ is the curvature scalar of the Weyl geometry. This is defined in terms of the affine connection of the Weyl space:

\bea \Gamma^\mu_{\alpha\beta}=\{^\mu_{\alpha\beta}\}+\frac{1}{2}\left(\delta^\mu_\alpha w_\beta+\delta^\mu_\beta w_\alpha-g_{\alpha\beta} w^\mu\right),\label{affine-c}\eea where $\{^\mu_{\alpha\beta}\}$ are the standard Christoffel symbols of the metric (properly the affine connection of the Riemann space). The gauge covariant derivative of the Higgs field is defined as $D_\mu h:=(D^*_\mu-w_\mu/2)h$, where $D^*_\mu h$ is the gauge covariant derivative in the standard EW theory. It is evident that, if set the mass parameter $v_0=0$, the action (\ref{cheng-action}) is invariant under the Weyl rescalings (\ref{weyl-r}) plus the scale transformation:

\bea h\rightarrow \Omega^{-1} h\;(\Rightarrow\,D_\mu h\rightarrow\Omega^{-1}D_\mu h).\label{scale-t-h}\eea However, given that, in general, the mass parameter is non-vanishing ($v^2_0\neq 0$), then the EW symmetry breaking potential in (\ref{cheng-action}) not only allows for generation of masses of the gauge bosons (and fermions) but, also, generates the Planck mass $M_\text{pl}=\sqrt\xi\,v_0$, where $v_0\approx 246$ GeV, and $\xi\sim 10^{32}-10^{34}$ is too large to meet the observational constraints.\footnote{The first bound on the value of the non minimal coupling ($\xi<2.6\times 10^{15}$) was derived in \cite{calmet}.} Hence, the non vanishing mass parameter $v_0$ not only allows for EW symmetry breaking but, also, breaks the scale invariance of the theory. In the general relativity (GR) gauge where $$w_\mu=0\;\Rightarrow\;R^{(w)}\rightarrow R,\;D_\mu h\rightarrow D^*_\mu h,$$ the theory (\ref{cheng-action}) is known as ``induced gravity'' \cite{ind-grav}. Among other things this theory has been investigated as a model of Higgs inflation \cite{cervantes-cota}. A trivial modification of the induced gravity model which is given by the action:

\bea &&S=\int d^4x\sqrt{|g|}\left[\frac{M^2_\text{pl}+\xi|h|^2}{2}\,R-\right.\nonumber\\
&&\left.\;\;\;\;\;\;\;\;\;\;\;\;\;\;\;g^{\mu\nu}(D^*_\mu h)^\dag(D^*_\nu h)-\frac{\lambda}{4}(|h|^2-v_0^2)^2\right],\label{bezrukov-action}\eea where all quantities are in the GR gauge ($w_\mu=0$), has been investigated in \cite{bezrukov} as a model to address both inflation and particle phenomenology,\footnote{In \cite{salvio} a bound on the mass of the Higgs field in the model (\ref{bezrukov-action}) was derived, revealing some tension with the experiments.} while in \cite{rinaldi} it has been shown to provide a description of both primordial inflation and late-time acceleration of the cosmic expansion (as well as the dark matter in the universe), if the dynamics of the Higgs field's phase is taken into account. A question then arises: can be a model like (\ref{bezrukov-action}) compatible with scale invariance? If yes, how to modify this theory in order to achieve scale invariance?

\subsection*{Scale invariant theory}

To answer the above questions let us come back to Weyl geometry. It is known since long ago that a drawback of Weyl geometric theory is associated with non-integrability of length in this theory: under parallel transport of a vector along a closed path in spacetime its length is changed according to $l=l_0\exp\oint dx^\mu w_\mu/2$. This might be associated with an unobserved broadening of the atomic spectral lines, also known as the ``second clock effect'' \cite{novello}. There is a simpler variant of Weyl geometry called as ``Weyl integrable geometry'' (WIG), which is free of the mentioned problem. WIG is obtained from Weyl theory if make the replacement $w_\mu\rightarrow\der_\mu\vphi$, where $\vphi$ is known as the Weyl gauge scalar. In this case, since $\oint dx^\mu \der_\mu\vphi/2=0$, the length of a vector is integrable. Although several authors consider the above replacement as a trivial gauge and identify the resulting geometry with standard Riemann space, this is not true. In fact, in the obtained affine structure, the (integrable) lengths of vectors are actually point dependent. As a result, the affine connection (\ref{affine-c}) with the replacement $w_\mu\rightarrow\der_\mu\vphi$, the non-metricity condition of WIG, the corresponding WIG Riemann-Christoffel and Ricci tensors, and the covariant derivative operator\footnote{From this point on all quantities labeled with the ``$(w)$'' refer to Weyl integrable objects which are defined with respect to the affine connection (\ref{affine-c}) after the replacement $w_\mu\rightarrow\der_\mu\vphi$.} $$\;\nabla^{(w)}_\mu g_{\alpha\beta}=-\der_\mu\vphi\,g_{\alpha\beta},\;R^{(w)}_{\alpha\beta\mu\nu},\;R^{(w)}_{\mu\nu},\;\nabla^{(w)}_\mu,$$ among other quantities, are all invariant under the following Weyl rescalings:

\bea g_{\mu\nu}\rightarrow\Omega^{2}g_{\mu\nu},\;\vphi\rightarrow\vphi-2\ln\Omega.\label{scale-t}\eea This means that there is not a single WIG spacetime $({\cal M},g_{\mu\nu},\vphi)$, but a whole equivalence class of them: ${\cal C}=\{({\cal M},g_{\mu\nu},\vphi):\nabla^{(w)}_\mu g_{\alpha\beta}=-\der_\mu\vphi\,g_{\alpha\beta}\}$, such that any other pair $(\bar g_{\mu\nu},\bar\vphi)$ related with $(g_{\mu\nu},\vphi)$ by (\ref{scale-t}), also belongs in the conformal equivalence class ${\cal C}$. This property is not shared by Riemann geometry which corresponds to the particular GR gauge: $\vphi=\vphi_0=const$.

The minimal gravitational action associated with WIG background which is invariant under the local scale transformations (\ref{scale-t}) is $\int d^4x\sqrt{|g|}M^2_\text{pl}e^\vphi R^{(w)}/2$. Since the effective Planck mass $M^2_\text{pl}(\vphi)=M^2_\text{pl}\,e^\vphi$ is a point dependent quantity, then, assuming $M_\text{pl}(\vphi)$ to be the standard unit of mass, any mass parameter should share the same point dependent property: $v_0^2\rightarrow v_0^2(\vphi)=v^2_0\,e^\vphi$. The above is a direct consequence of adopting WIG backgrounds where the lenght of any vectors is point dependent, as the geometrical arena for the gravitational phenomena. If one wants to make contact with the SMP, the terms within the action (\ref{cheng-action}) should be added to the above discussed minimal scale invariant action. The resulting scale invariant WIG-SMP action is

\bea &&S=\int d^4x\sqrt{|g|}\left[\frac{M^2_\text{pl}(\vphi)+\xi|h|^2}{2}\,R^{(w)}-\right.\nonumber\\
&&\left.\;\;\;\;\;\;\;\;\;\;\;\;\;g^{\mu\nu}(D_\mu h)^\dag(D_\nu h)-\frac{\lambda}{4}(|h|^2-v_0^2(\vphi))^2\right],\label{s-i-action}\eea where -- we recall -- the gauge covariant derivative of the Higgs field is defined as $D_\mu h=(D^*_\mu-\der_\mu\vphi/2) h$, and

\bea M^2_\text{pl}(\vphi)\equiv M^2_\text{pl}\,e^\vphi,\;v^2_0(\vphi)\equiv v_0^2\,e^\vphi.\label{m-v0}\eea This action differs from the one in \cite{shaposhnikov} -- see also \cite{indios} -- in that the underlying geometric structure is WIG, and $\vphi$ is no longer another singlet scalar field but it is just the Weyl gauge field of WIG geometry, i. e., the $\vphi$ kinetic energy term is already included in the WIG curvature scalar $R^{(w)}$. Since under the Weyl rescalings (\ref{scale-t}): $M_\text{pl}(\vphi)\rightarrow\Omega^{-1}M_\text{pl}(\vphi)$, $v_0(\vphi)\rightarrow\Omega^{-1}v_0(\vphi)$, it is a simple exercise to show that the action (\ref{s-i-action}) with the definitions (\ref{m-v0}) is invariant under the local scale transformations (\ref{scale-t}), (\ref{scale-t-h}). In the GR gauge the theory (\ref{bezrukov-action}) is recovered. 

For simplicity in (\ref{s-i-action}) we have omitted the EW Lagrangian terms but for the Higgs boson. However, it has been demonstrated that the EW terms missing in (\ref{s-i-action}) do not spoil the scale invariance in Weyl spaces \cite{cheng}. Hence, the action (\ref{s-i-action}) is not only scale invariant but, also, it perfectly accommodates the SMP. In this theory the gravity is propagated both by the metric and by the gauge Weyl scalar, so that this is a scalar-tensor theory. However, unlike other scalar-tensor theories like, for instance, Brans-Dicke gravity \cite{brans}, since both $g_{\mu\nu}$ and $\vphi$ contribute towards the curvature of spacetime, in the present theory gravity is a fully geometrical phenomenon. In consequence, since the mass of bosons and fermions of the SMP are related with the point dependent symmetry breaking mass parameter $v_0(\vphi)$, the masses of the elementary particles are influenced by the spacetime curvature through the gauge scalar $\vphi$.\footnote{Here we assume that not only the mass of elementary particles, but also the masses $m_\textsc{s}$ of composite systems like hadrons, atoms, molecules, etc., and of macroscopic bodies, depend on spacetime point as $v_0(\vphi)$: $m_\textsc{s}=m_\textsc{s,0}\,e^{\vphi/2}$. This assumption is consistent with experimental evidence on the equivalence principle \cite{will} and on variation of electron-to-proton mass relation \cite{sudarsky}.}

One interesting property of the theory (\ref{s-i-action}) is a direct consequence of scale invariance: There is a whole conformal equivalence class ${\cal C}$ of spacetimes, which amount to equivalent geometrical descriptions of a same phenomenon. Mathematically this means that, in addition to the four degrees of freedom to make spacetime diffeomorphisms, a new degree of freedom to make scale transformations arises. To illustrate this let us consider vacuum cosmology within the context of the theory (\ref{s-i-action}). Assume a Friedmann-Robertson-Walker (FRW) spacetime with flat spatial sections, given by the line element $ds^2=-dt^2+a^2(t)\delta_{ij}dx^idx^j$, where $t$ is the cosmic time and $a(t)$ is the scale factor. The vacuum field equations derived from (\ref{s-i-action}): $G^{(w)}_{\mu\nu}\equiv R^{(w)}_{\mu\nu}-g_{\mu\nu}R^{(w)}/2=0$, can then be written as follows: $3(H+\dot\vphi/2)^2=0$, $\dot H+\ddot\vphi/2=0$, where $H\equiv\dot a/a$, and the dot accounts for $t$-derivative. It is to be remarked that the Klein-Gordon (KG) equation is not an independent equation but it is just the trace of the Einstein's equations. The Friedmann equation can be integrated to obtain the following dependence of the scale factor upon the gauge field $\vphi$: $a(\vphi)\propto e^{-\vphi/2}$. If we substitute this $a(\vphi)$ back into the remaining equations -- second field equation above and the KG equation -- these become just identities so that no new information can be extracted from them. This is a consequence of scale invariance since we have the freedom to choose either any $\vphi(t)$ or any $a(t)$ we want. Recall that one of these degrees of freedom can be transformed in any desired way by an appropriate scale transformation (\ref{scale-t}).

\subsection*{Measurable quantities}

In the presence of matter the WIG-Einstein equations which are obtained from (\ref{s-i-action}) read: 

\bea G^{(w)}_{\mu\nu}=T^{(m)}_{\mu\nu}/M^2_\text{pl}(\vphi)=e^{-\vphi}T^{(m)}_{\mu\nu}/M^2_\text{pl}=T^{(m,w)}_{\mu\nu}/M^2_\text{pl},\label{e-w-feq}\eea where $T^{(m)}_{\mu\nu}$ is the usual stress-energy tensor of matter, while $T^{(m,w)}_{\nu\mu}\equiv e^{-\vphi}T^{(m)}_{\mu\nu}$ is the corresponding WIG tensor. The Bianchi identity $\nabla^\nu_{(w)} G^{(w)}_{\nu\mu}=0$ entails the following conservation equation: $\nabla^\nu_{(w)}T^{(m,w)}_{\nu\mu}=0$, meaning that it is the WIG stress-energy tensor $T^{(m,w)}_{\nu\mu}$ the one which is conserved in WIG spacetimes. Besides, if adopt an extended version of the principle of general covariance according to which the equations of physics should have tensorial form, with the involved tensors transforming in the same way under general coordinate transformations and Weyl rescalings (\ref{scale-t}), it is $T^{(m,w)}_{\nu\mu}$ and not $T^{(m)}_{\mu\nu}$ which has the physical meaning. Notice that both tensors involved in the WIG-Einstein equations (\ref{e-w-feq}): $T^{(m,w)}_{\nu\mu}$ and $G^{(w)}_{\mu\nu}$, transform in the same way under (\ref{scale-t}) (as a matter of fact these are not transformed by (\ref{scale-t})). The components of the WIG stress-energy tensor $T^{(m,w)}_{00}$, $T^{(m,w)}_{0i}$, $T^{(m,w)}_{ij}$, correspond to the physical parameters which are measured in experiments performed in WIG backgrounds. In the Newtonian limit where particles move slowly $dx^i/ds\ll dt/ds$, and only weak static gravitational field is considered: $g_{\mu\nu}=\eta_{\mu\nu}+h_{\mu\nu}$, $\vphi=\vphi_0+\phi$ ($|h_{\mu\nu}|\ll 1$, $\phi\ll 1$, $\der h_{\mu\nu}/\der t=0$, $\der\phi/\der t=0$), one has that $d^2x^i/dt^2=\der_i (h_{00}-\phi)/2$ $\Rightarrow\;h_{00}-\phi=-2\Phi_\textsc{n}$, where $\Phi_\textsc{n}$ is the Newtonian gravitational potential. In the same limit, since $T^{(m,w)}_{00}\gg|T^{(m,w)}_{ik}|$, $T^{(m,w)}=-T^{(m,w)}_{00}$, the WIG-Einstein's field equations lead to ($\nabla^2\equiv \der^2_i$): $$R^{(w)}_{00}=(1/M^2_\text{pl})T^{(m,w)}_{00}\;\Rightarrow\;\nabla^2\Phi_\textsc{n}=T^{(m,w)}_{00}/2M^2_\text{pl},$$ so that, for a homogeneous distribution of matter within the volume $\Omega$: $$\Phi_\textsc{n}=-G_\textsc{n}\frac{M}{r},\;M=\int_\Omega d^3x\,T^{(m,w)}_{00},$$ where it is apparent that it is the constant $G_\textsc{n}=M^{-2}_\text{pl}/8\pi$ the one which is is measured in Cavendish experiments.

Another meaningful quantity is the four-momentum $p_\mu=g_{\mu\nu}p^\nu$ ($p^\mu:=m_0 e^{\vphi/2} dx^\mu/d\tau$, $d\tau=-i ds$). Since, under (\ref{scale-t}) the gauge boson $A_\mu$ is unchanged -- as well as $p_\mu$ -- the above definition of the momentum admits a gauge extension which is consistent with scale-invariance: $p_\mu\rightarrow p_\mu+e A_\mu$ $\Leftrightarrow\;\der_\mu\rightarrow\der_\mu+ie A_\mu$. The geodesic equation of a particle with momentum $p^\mu$ moving in a WIG background is given by:

\bea \frac{dp^\mu}{ds}+\Gamma^\mu_{\;\sigma\lambda} \frac{dx^\sigma}{ds}\,p^\lambda=\der_\lambda\vphi\frac{dx^\lambda}{ds}\,p^\mu.\label{geod-eq}\eea The term in the RHS of this equation is associated with the point-dependent property of the particle's mass in WIG spacetimes and has nothing to do with any additional force. This is evident if, after an appropriate affine parametrization $ds\rightarrow d\sigma=e^{\vphi/2}ds$, rewrite the Eq. (\ref{geod-eq}) in the equivalent form: $$\frac{d}{d\sigma}\left(\frac{dx^\mu}{d\sigma}\right)+\Gamma^\mu_{\;\nu\lambda} \frac{dx^\nu}{d\sigma}\frac{dx^\lambda}{d\sigma}=0.$$

\subsection*{Mass hierarchy}

A nice result of the present study is that nothing besides the scale invariant theory (\ref{s-i-action}) and inflation is required to explain the large hierarchy between the Higgs and Plack masses. Actually, suppose initial conditions are given in the neighborhood of the local maximum of the EW symmetry breaking potential $V(|h|)=\lambda(|h|^2-v^2_0(\vphi))^2/4$, i. e., at $h=0$. Let us assume, for simplicity, the following initial conditions: $\vphi(0)=0$, $M^2_\text{pl}(0)=M^2_\text{pl}$, $v_0^2(0)=v_0^2$. The resulting theory is just Einstein-Hilbert (de Sitter) gravity: $S=M^2_\text{pl}\int d^4x\sqrt{|g|}(R-\lambda v_0^4/2M^2_\text{pl})/2.$ Since at $h=0$ the EW symmetry is unbroken, the particles of the SMP remain massless. Suppose, besides, that the initial value of the mass parameter is of the order of the Plack scale ($v_0\sim 10^{19}$ GeV), so that, assuming $\lambda\approx 1$, a large cosmological constant $\lambda v_0^4/4M^2_\text{pl}$ may fuel primordial inflation. After that the universe starts inflating and the dynamics is described by the action (\ref{s-i-action}). 

In order to estimate the impact of inflationary expansion on the masses of SMP particles, let us assume that during the inflationary period the gravitational dynamics is approximately dictated by the vacuum Einstein-Weyl equations $G^{(w)}_{\mu\nu}=-\Lambda\,e^\vphi g_{\mu\nu}$ $\Rightarrow\;3(H+\dot\vphi/2)^2=\Lambda\,e^\vphi$, where $\Lambda\equiv\lambda v^4_0/4M^2_\text{pl}$. If assume de Sitter expansion [$H=H_0\Rightarrow a=a_0\exp(H_0\,t)$], then the relationship between the scale factor and the gauge scalar is given by: $e^{\vphi/2}=1/(1+k\,a)$, where $k=1/a_0$ is an arbitrary constant. At small $k a\ll 1$, $\vphi$ is almost a constant, which corresponds to the GR gauge, while at large $k a\gg 1$, $e^{\vphi/2}\propto a(t)^{-1}\propto e^{-H_0 t}$, where for simplicity we assume that $H_0=\sqrt{\Lambda/3}$. By the end of inflation, once the Higgs field settles down in the minimum of the symmetry breaking potential, acquiring a point-dependent vev $h=v_0(\vphi)$, the SMP particles acquire masses: $m(\vphi)\propto v_0(\vphi)=v_0\,e^{\vphi/2}$, leaving no trace of the initially existing effective cosmological constant $\Lambda$. It immediately follows that inflation and point-dependence of mass are enough to explain the presently small mass of the SMP particles as compared with the Plack scale. Actually, suppose that during inflation the linear size of the universe has expanded by a factor $f=a_\text{fin}/a_\text{ini}$. This means that during the inflationary stage the mass parameter would have decreased by the inverse factor: $v^\text{fin}_0/v^\text{ini}_0=a_\text{ini}/a_\text{fin}=f^{-1}$. The masses of the SMP particles are set by the mass parameter $v^\text{fin}_0$: $m\propto v^\text{fin}_0\propto f^{-1}v_0$. Hence, a modest inflationary factor $f\sim 10^{16}$ is required to explain the large hierarchy between the Higgs mass $m_\textsc{h}\sim 1$ TeV and the Planck mass scale $v_0\sim M_\text{pl}\sim 10^{19}$ GeV. Since a factor of at least $10^{27}$ is required to explain all of the puzzles inflation solves \cite{liddle}, this means that we do not need the entire inflationary epoch but just its final stages where our assumptions are justified. To put the above explanation on a sound footing it suffices to point out that, as stated above, the actually measured value of the Cavendish gravitational constant is $G_\textsc{n}=M^{-2}_\text{pl}/8\pi$ (and not the very much larger $M^{-2}_\text{pl}(\vphi)/8\pi$), meanwhile the masses of individual particles and of composite systems are usually measured through methods involving their interaction with electromagnetic fields in the presence of background gravity as, for instance, in mass spectrometers. In this latter case it is the geodesic equation (\ref{geod-eq}) what matters, meaning that it is the varying mass $m(\vphi)=e^{\vphi/2} m_0$ which is measured in this case.

The author thanks Stanley Deser and Alberto Salvio by useful comments, and SNI of Mexico for support of this research.

\end{document}